\documentclass{pinchcr}   

\usepackage{amsmath}

\title{Quantum optics with diamond color centers coupled to nanophotonic devices}

\author{Alp Sipahigil and Mikhail D. Lukin}
\affiliation{Department of Physics, Harvard University, Cambridge, MA 02138 USA}

\begin{document}

\maketitle
\preface{
We review recent advances towards the realization of quantum networks based on atom-like solid-state quantum emitters coupled to nanophotonic devices. Specifically, we focus on experiments involving the negatively charged silicon-vacancy color center in diamond. These emitters combine homogeneous, coherent optical transitions and a long-lived electronic spin quantum memory. We discuss optical and spin properties of this system at cryogenic temperatures and describe experiments where silicon-vacancy centers  are coupled to nanophotonic devices.  Finally, we discuss experiments demonstrating  quantum nonlinearities at the single-photon level and two-emitter entanglement in a single nanophotonic device.
}
 
\tableofcontents
\maintext
\chapter{Quantum optics with diamond color centers coupled to nanophotonic devices}

\section{Introduction: Quantum optics with solid-state systems}
\label{Introduction}


In the past two decades major advances in the isolation and coherent control of individual quantum systems ranging from neutral atoms, ions, and single-photons to Josephson junction based circuits and electron spins in solids\,\shortcite{winelandNobel,harocheNobel} have been made. The ability to control single quantum systems has opened up new possibilities ranging from  quantum information processing and simulation\,\shortcite{cirac1995quantum,jaksch1998cold}, to quantum-enhanced sensing\,\shortcite{maze2008nanoscale,Rondin2014} and quantum communication\,\shortcite{Cirac1997,kimble2008quantum}. 
Many of these potential applications require the realization of controllable interactions between multiple qubits. Several techniques have been developed to create such interactions between qubits over short distances at the single device level. Examples include dipole-dipole interactions between neutral atoms\,\shortcite{jaksch2000fast,urban2009observation}, phonon-mediated interactions in trapped ion crystals\,\shortcite{cirac1995quantum,monroe1995demonstration} and photon-mediated interactions in cavity and circuit quantum electrodynamics (QED)\,\shortcite{pellizzari1995decoherence,neuzner2016interference,majer2007coupling}. Recent experiments have used these interactions to entangle up to a few tens of qubits \,\shortcite{Monz2011,barends2014superconducting} and control quantum dynamics of more then 50 qubits\,\shortcite{bernien2017,zhang2017}. The current challenge involves scaling these systems up to a large number of connected, controlled  qubits.  

One approach to scalability involves distributing quantum information between remote nodes that each contain multiple qubits\,\shortcite{kimble2008quantum}. This approach has motivated the development of quantum networks where stationary qubits that can store quantum information are interfaced with optical photons for distributing quantum information\,\shortcite{monroe2014large,monroe2016quantum,andrews2014bidirectional}. Such nodes also form the basis of quantum repeater architectures for long-distance quantum communication, where stationary qubits are used as quantum memories for optical photons\,\shortcite{Briegel1998}. 

Optical interfaces for different qubit architectures are now being actively explored. Microwave-to-optical frequency conversion using mechanical transducers is being investigated to interface microwave qubits based on Josephson junctions with optical photons\,\shortcite{andrews2014bidirectional}. 
For optical emitters such as neutral atoms and trapped ions, cavity QED techniques have been developed to interface hyperfine qubits with optical photons\,\shortcite{Cirac1997,Ritter2012}. 
In this chapter, we will discuss recent advances using color centers in diamond that demonstrate key elements required to realize a solid-state quantum network node. 

Color centers in diamond are part of a growing list of solid-state optical emitters\,\shortcite{aharonovich2016solid} that include quantum dots\,\shortcite{Michler2000}, rare-earth ions\,\shortcite{Lvovsky2009}, single molecules\,\shortcite{Orrit1990},  and point defects in crystals\,\shortcite{Gruber1997}. The recent interest in solid-state emitters for quantum applications builds upon a long history of laser-science, optoelectronics and microscopy  research. Since 1960s, transition-metal and rare-earth ions embedded in solids have found wide use as an optical gain medium\,\shortcite{maiman1960stimulated} based on their weakly-allowed optical transitions\,\shortcite{judd1962optical,liu2006spectroscopic}. These studies enabled applications ranging from the development of optical amplifiers based on erbium-doped silica fibers which form the basis of modern telecommunication infrastructure\,\shortcite{mears1987low} to widely tunable solid-state lasers such as Ti:Sapphire lasers. 

The early investigations of  solid-state emitters were primarily interested in understanding the properties and applications of ensembles of dopants and point defects\,\shortcite{liu2006spectroscopic,stoneham1975theory}. The desired properties for quantum applications, however, differ strongly from those of a gain medium. Specifically, the ability to isolate and control single emitters is critical for realizing controllable qubits. Moreover, in order to achieve coherent atom-photon interactions based on single solid-state emitters, it is necessary to identify systems with strong transition dipole moments (to enable high radiative decay rates), weak static dipole moments (to minimize optical dephasing due to environmental noise), and weak vibronic coupling (to minimize phonon broadening). In addition, the presence of metastable spin sublevels is necessary to create a long-lived memory. 
 
By early 1990s, advances in optical spectroscopy and microscopy enabled the first major step in this direction, the optical detection of single molecules inside solid-state matrices\,\shortcite{Moerner1989,Orrit1990}. In the past two decades, these techniques have been extended to studies of quantum dots and nitrogen-vacancy (NV) color centers in diamond. These systems have shown promising spin and optical properties which made them leading candidates for use in solid-state quantum network nodes\,\shortcite{gao2015coherent}. 

Optically active quantum dots are mesoscopic semiconductor structures where electrons and holes are confined to result in a discrete, atom-like optical spectra\,\shortcite{Michler2000,Santori2002}. Self-assembled InGaAs quantum dots embedded in a GaAs matrix have strong optical dipole transitions with excited state lifetimes in the range of $\sim 1$~ns. They can be integrated into nanophotonic structures\,\shortcite{lodahl2015interfacing} to achieve strong light-matter interactions at high bandwidths\,\shortcite{Hennessy2007,englund2007controlling}. However, the spatial and spectral positions of these emitters are non-deterministic due to the self-assembly process\,\shortcite{lodahl2015interfacing}. 

Charged quantum dots also have spin degrees of freedom which can be used to store quantum information. The coherence of the optical quantum dot spin is limited by the high density nuclear spin bath in the host crystal which leads to an inhomogeneous spin dephasing timescale of $T_2^*$ of $\sim2$~ns and a coherence time ($T_2$) of 1-3~$\mu\textrm{s}$ using dynamical decoupling sequences\,\shortcite{Greilich2006,Press2010,Bechtold2015,Stockill2016}. The few microsecond long coherence time limits the potential use of InGaAs quantum dots as quantum memories for long-distance quantum communication applications. 

The nitrogen-vacancy (NV) color center in diamond consists of a substitutional nitrogen atom with a neighboring vacancy\,\shortcite{Gruber1997,Doherty2013}. The NV center has a spin-triplet ground state with a long coherence time of up to $2$~ms at room temperature\,\shortcite{balasubramanian2009ultralong}. Interestingly, the NV spin can be optically polarized and read out at room temperature using off-resonant excitation\,\shortcite{Jelezko2004}. The ability to initialize, coherently control and read out the NV spin at room temperature using a simple experimental setup resulted in its wide use in nanoscale sensing applications such as nanoscale magnetometery and thermometry\,\shortcite{Taylor2008,lovchinsky2016,Kucsko2013}. 
At cryogenic temperatures, about $4\%$ of the NV fluorescence is emitted into a narrowband zero-phonon line (ZPL). For NV centers in bulk diamond, the linewidth of this ZPL transition can be close to the lifetime-limited linewidth of $13~$MHz where the excited state lifetime is $12$\,ns\,\shortcite{tamarat2006stark}. For NV centers  that are close to surfaces (e.g. in a nanostructure), electric field noise due to charge fluctuations in the environment results in spectral diffusion of the optical transitions and non-radiatively broadens the optical transition to few GHz\,\shortcite{faraon2012coupling,Riedel2017}. This poses a major challenge for realizing a nearly-deterministic spin-photon interface based on an NV center in a microphotonic device. In the past decade, efforts to overcome these challenges focused on integrating NV centers in photonic crystals\,\shortcite{Englund2010,faraon2012coupling,Hausmann2013} or fiber-based Fabry-Perot cavities\,\shortcite{Albrecht2013,Riedel2017} for Purcell-enhancing the ZPL emission rate. In addition, methods to minimize the electric field noise originating from surfaces and fabrication-induced damage are being actively explored\shortcite{chu2014coherent}. 

In this Chapter, we discuss recent progress in addressing these challenges using a new family of color centers in diamond coupled to nanophotonic devices. Specifically, we focus on the optical and spin coherence properties of  the silicon-vacancy (SiV) center\,\shortcite{neu2011single,HeppThesis,hepp2014electronic} at cryogenic temperatures that are relevant for applications in quantum science and technology. This solid-state platform, which combines coherent optical and spin transitions, recently emerged as a promising approach for the realization of scalable quantum networks.  

This Chapter is organized as follows. 
In Section~\ref{coherentoptics} we discuss the relevant broadening mechanisms of the optical transitions for solid-state emitters and the figures of merit to achieve a deterministic spin-photon interface. In Section~\ref{indistinguishable}, we show that SiV centers in high-quality diamond crystals have a narrow inhomogeneous distribution and can efficiently generate indistinguishable photons. Section~\ref{SiVimplanted} shows that the optical coherence properties of SiV centers are maintained in nanophotonic structures. Section~\ref{SiVnanophotonics} discusses experiments where SiV centers are strongly coupled to photonic crystal cavities to achieve a deterministic spin-photon interface and optical nonlinearities at the single photon level.  Entanglement generation between two SiV centers in a single nanophotonic device is described in Section~\ref{twoSiVentanglement}. The spin properties of the SiV center which are important for realizing a long-lived quantum memory are discussed in Section~\ref{coherencesection}. This section presents a microscopic model of the electron-phonon interactions that result in spin dephasing and recent experiments that obtained $13$\,ms coherence at dilution fridge temperatures. We conclude with an outlook discussing prospects for realizing quantum repeater nodes based on long-lived SiV spin qubits strongly interacting with optical photons.

\section{Coherent atom-photon interactions using solid-state emitters}
\label{coherentoptics}

Solid-state quantum emitters are tightly confined in a crystal matrix. This allows the  placement of emitters in the close vicinity of  dielectric or plasmonic nanophotonic structures without the challenges associated with trapping neutral atoms or ions near surfaces. The ability to position emitters in the near field of strongly confined modes enables strongly enhanced light-matter interactions. Using this approach, Purcell enhancement of optical transitions~\shortcite{Purcell1946,lodahl2004controlling} has been demonstrated for a wide range of quantum emitters\,\shortcite{aharonovich2016solid}. Strong coupling has also been achieved using quantum dots embedded in photonic crystal cavities\,\shortcite{Senellart2017}. 

Purcell-enhanced emission into nanophotonic modes ensures that the emitters interact with a well-defined spatial mode as opposed to $4\pi$ emission into free space. However, for most applications in quantum information science, the temporal mode also needs to be well defined. In other words, the emission linewidth ($\gamma$) should match the lifetime limit ($\gamma_{rad}=1/T_1$) to be able to generate indistinguishable single photons\,\shortcite{Lettow2010,Legero2003}. While this is often true for cold atomic systems with radiatively broadened transitions ($\gamma=\gamma_{rad}$), the complex environment of solid-state emitters can result in homogeneous and inhomogeneous dephasing processes ($\gamma_d$) that nonradiatively broaden the optical transitions with $\gamma=\gamma_{rad}+\gamma_d$.

Phonons and strain can provide such additional homogeneous and inhomogeneous broadening mechanisms.
Displacements of atoms in the host crystal can affect the optical transitions in two different ways.  
Static lattice distortions (i.e. strain) may reduce the symmetry of the defect and change the energy splittings\,\shortcite{sternschulte1994luminescence} between electronic orbitals.
A variation in local strain contributes to the inhomogeneous distribution of the resonance frequencies.  
Displacements of the lattice can also give rise to dynamic effects during an optical excitation cycle.  A homogeneous dephasing mechanism is caused by elastic and inelastic scattering of bulk acoustic phonons. While this process does not modify the single-photon generation abilities at room temperature, it limits the coherence time of the generated photons typically below a ps timescale. Photons generated at room temperature are therefore not suitable for the observation of quantum interference effects. The dephasing due to acoustic phonons can often be completely suppressed by operating at liquid helium temperatures. A microscopic model of this acoustic phonon broadening mechanism is discussed in Refs.\,\shortcite{fu2009observation,goldman2015phonon} for NV centers and in  Ref.\,\shortcite{jahnke2015electron} for SiV centers. 

\begin{figure}
\begin{center}
		\includegraphics[width=.8\linewidth]{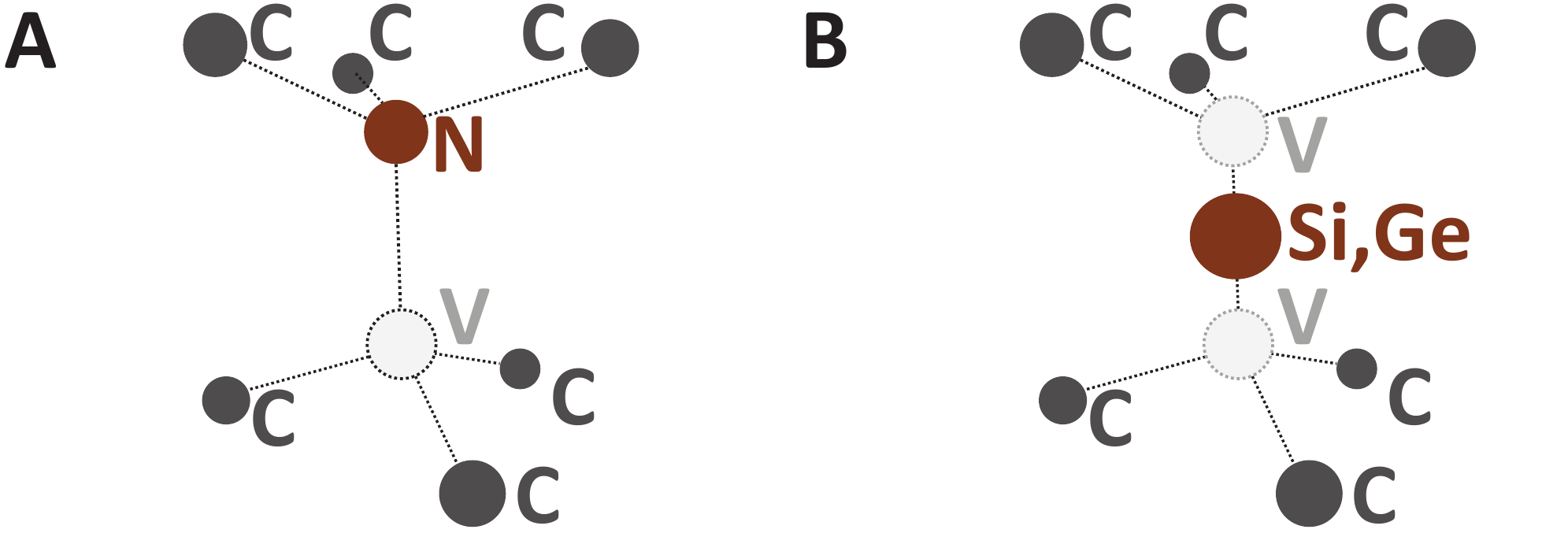}
\end{center}
\caption{Molecular structure of the NV, SiV and GeV centers in diamond. (A) The NV center consists of a substitutional nitrogen (N) with a neighboring lattice vacancy (V). (B) For the SiV and GeV centers, the impurity atom (Si or Ge) is centered between two lattice vacancies (V)  and constitutes an inversion center.}
\label{fig:nvsiv}
\end{figure}

In addition, at both ambient and cryogenic temperatures, the emission of a photon can be accompanied by the spontaneous creation of local, high-frequency (few THz), short-lived (ps decay timescale) molecular vibrations of the color center based on the Franck-Condon principle. Since these high-frequency localized modes decay very rapidly into bulk phonon modes\,\shortcite{Huxter2013}, only the zero-phonon line (ZPL) emission has sufficiently narrow linewidths for indistinguishable photon generation. The probability of emission into the ZPL is $\sim4\%$ for the NV center and $\sim70\%$ for the SiV center. The high emission probability into the ZPL for the SiV center can be understood based on the following consideration. 
%
%
The SiV optical transitions take place between orbital states of different parity, ${}^2\textrm{E}_\textrm{g}$ and ${}^2\textrm{E}_\textrm{u}$ , which differ in phase but have similar charge densities~\shortcite{Gali2013}.
This small change in the electronic charge density results in the strong ZPL, since optical excitations do not couple efficiently to local vibrations.

For NV centers, the dominant dephasing process at cryogenic temperatures is caused by charge fluctuations in the solid-state environment. As shown in Fig. \ref{fig:nvsiv}, the NV center consists of a nitrogen impurity and a neighboring lattice vacancy each occupying a substitutional site. This structure resembles that of a polar molecule and results in electronic orbitals with a large static dipole moment ($\sim 1$\, Debye). The energy of these electronic orbitals therefore shifts with any applied electric field, resulting in a broadening of the optical transition with charge fluctuations in the environment. The level of charge noise and spectral diffusion can be minimal for NV centers in high-quality single crystal bulk crystals\,\shortcite{Bernien2012,sipahigil2012quantum}. However, this dephasing process becomes more severe inside nanofabricated structures due to increased exposure to surface and defect states\,\shortcite{faraon2012coupling}. 

We recently demonstrated that silicon-vacancy (SiV)\,\shortcite{evans2016narrow,Sipahigil2016} and germanium-vacancy (GeV)\,\shortcite{bhaskar2017} centers in diamond nanophotonic structures maintain their optical coherence with a strong supression of the spectral diffusion observed in NV centers. The origin of this spectral stability for SiV and GeVs can be understood as a consequence of their inversion symmetry\,\shortcite{sipahigil2014indistinguishable}.  For defects such as SiV and GeV centers (Fig.~\ref{fig:nvsiv}) the impurity atom (Si or Ge) is located  at an interstitial site between two lattice vacancies. In this geometry, the impurity atom constitutes an inversion center. Owing to the inversion symmetry of the structure, electronic orbitals are symmetric with respect to the origin and have zero static electric dipole moments. The absence of a static electric dipole moment results in a first-order insensitivity to electric fields for SiVs and GeVs\footnote{We note that this property parallels a similar idea based on a charge-insensitive superconducting qubit design (i.e. the transmon qubit) which lead to major advances in circuit quantum electrodynamics in the past decade\,\shortcite{koch2007}.}.  This allows the incorporation of optical emitters with coherent optical transitions into nanophotonic devices which is the key enabling aspect for the experiments discussed in the following sections. 

\section{Indistinguishable photons from separated silicon-vacancy centers in diamond}
\label{indistinguishable}

The symmetry properties discussed in Section\,\ref{coherentoptics} lead to the absence of spectral diffusion\,\shortcite{rogers2014multiple} and a narrow inhomogeneous distribution\,\shortcite{sternschulte1994luminescence} for SiV centers in bulk diamond.
In Ref.~\shortcite{sipahigil2014indistinguishable}, we used SiV centers incorporated in a high-quality single-crystal diamond sample to generate indistinguishable photons from separate emitters. These SiV centers were incorporated in the high-quality crystals during the growth process and displayed a narrow inhomogeneous distribution of $\gamma_{inh}/2\pi\sim 1$\,GHz. For comparison, the natural linewidth of an SiV center is $\gamma_{nat}/2\pi=94$\,MHz corresponding to an excited state lifetime of $1.73$\,ns.  To our knowledge, this  ratio of the ensemble inhomogeneous distribution and the natural linewidth $\gamma_{inh}/\gamma_{nat}$ is the smallest observed among solid-state quantum emitters\,\shortcite{aharonovich2016solid}. This narrow distribution makes SiV centers suitable for indistinguishable photon generation from separated emitters.

\begin{figure}
\begin{center}
\includegraphics[width=\linewidth]{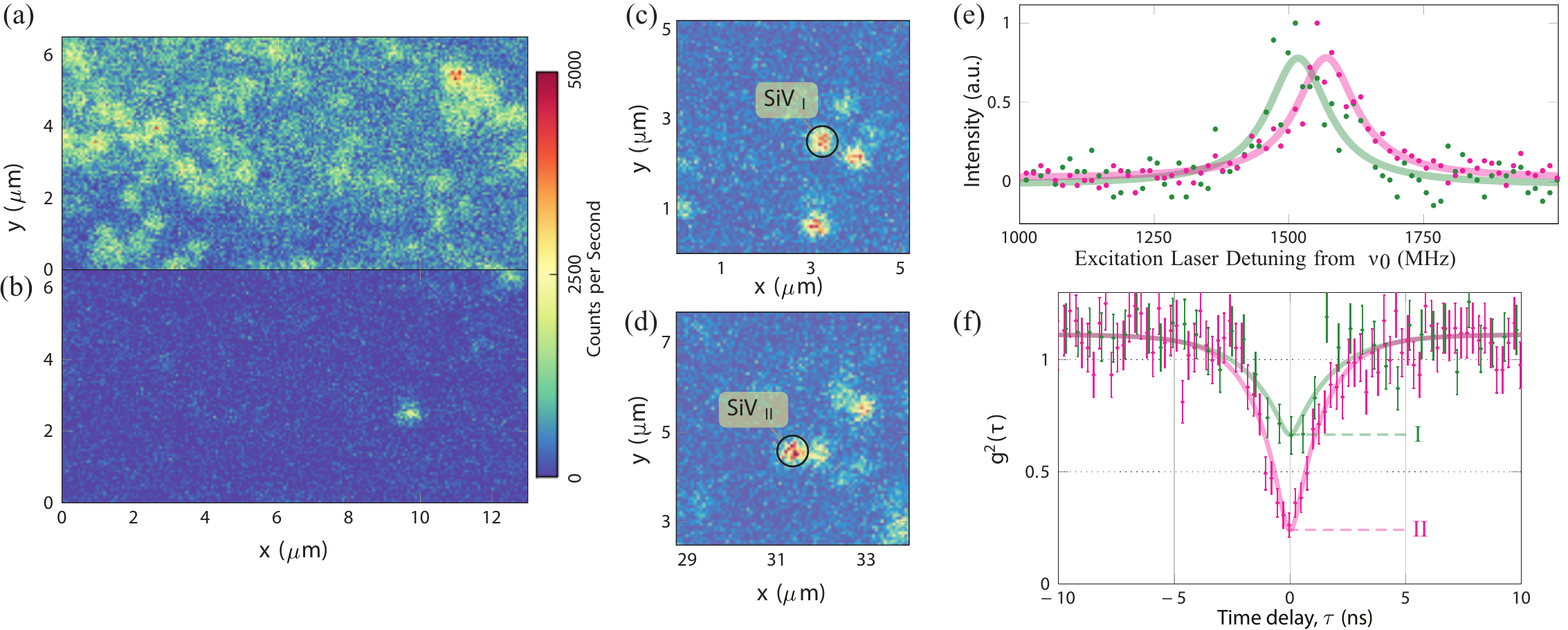}
\caption{(a) The probe laser frequency was fixed to the ensemble average of $\nu_0=406.7001$\,THz while spatially scanning the sample and monitoring fluorescence.  
	A high density of resonant emitters is visible with a large background.
	(b) Scan of the same region with the laser tuned to $\nu_1=\nu_0+1.5$\,GHz.  
	Due to the narrow inhomogeneous distribution, only few resonant emitters are visible and the background level is reduced. 
	(c) and (d) show the two emitters, SiV${}_\textrm{I}$ and SiV${}_\textrm{II}$, used for the two-photon interference experiment at frequency $\nu\sim\nu_1$. 
(e) PLE spectrum of the two SiVs: transition linewidth is $135\pm2$\,MHz for each emitter and the detuning is $52$\,MHz.  
(f) The single-photons from the two SiVs are interfered on a beamsplitter. The resulting second order intensity correlation function $g^2( \tau )$ at the output ports is plotted for two cases: 
(I) Single photons from the two emitters are chosen to be orthogonally polarized and hence distinguishable, $g^2_\perp (0) = 0.66 \pm 0.08$.
(II) For indistinguishable single photons with identical polarizations, $g^2_\parallel (0) = 0.26 \pm 0.05$. Figure adapted from (Sipahigil et al., 2014).
}
\label{fig:HOM}
\end{center}
\end{figure}

Figs.\ref{fig:HOM}(a) and (b) illustrate the narrow inhomogeneous distribution of the SiV centers by showing a  comparison of the density of SiV centers at the center frequency ($\nu_0$) of the inhomogeneous distribution and at a detuning of $1.5$ GHz. When we excite the sample at the center frequency of the ensemble $\nu_0$, a high density of emitters are visible with a large backround due to emitters in a different focal plane. Detuning the excitation laser only by $1.5$\,GHz from this frequency results in a drastic reduction in the density of visible emitters. 

To generate single photons from isolated, single SiV centers and minimize background from other emitters\,\shortcite{Moerner1989}, the laser was tuned to the edge of the inhomogeneous distribution ($\nu=\nu_0+1.5$\,GHz) where nearly-resonant, single SiV centers can be spatially resolved as shown in Figs.~\ref{indistinguishable}(c,d).  Photoluminescence excitation (PLE) spectra of the emitters, $\textrm{SiV}_\textrm{I}$ and $\textrm{SiV}_\textrm{II}$ ,  reveal transitions separated by 52.1\,MHz with a linewidth of 136 and 135\,MHz respectively.  For comparison, the lifetime of the excited states was measured to be $1.73\pm0.05$\,ns at temperatures below 50\,K corresponding to a transform limited linewidth of 94\,MHz. The narrow linewidths of the optical transitions are close to the lifetime-limited linewidth, indicating coherent single-photon generation suitable for efficient indistinguishable photon generation. 

To test the indistinguishability of the generated photons, single photons from $\textrm{SiV}_\textrm{I}$  and $\textrm{SiV}_\textrm{II}$ (Fig.\,\ref{fig:HOM}\,(c,d)) were directed to the input ports 1 and 2 of a beamsplitter respectively.  
Figure \ref{fig:HOM} shows two-photon interference\,\shortcite{Hong1987} measurements where the degree of indistinguishability of single photons is varied by changing the photon polarization. 
The two datasets show the second order intensity correlation function, $g^{2}(\tau)$, measured for indistinguishable (II) and distinguishable (I) photon states.  
For identically polarized indistinguishable photons, we find $g^2_\parallel (0) = 0.26 \pm 0.05$ where the error bars denote shot noise estimates.  
After rotating the fluorescence polarization of $\textrm{SiV}_\textrm{II}$  by $90^\circ$ to make the photon sources distinguishable, $g^2_\perp (0) = 0.66 \pm 0.08$ was observed.  
These results clearly demonstrate two-photon interference corresponding to a measured two-photon interference visibility of $\eta= 0.72 \pm 0.05$ that is limited by detector timing resolution and background photons\,\shortcite{sipahigil2014indistinguishable}.  

These observations established the SiV center as an excellent source of indistinguishable single photons with the combination of a strong ZPL transition, narrow inhomogeneous distribution, and spectrally stable optical transitions. 
These results also suggested the possibility of integrating SiV centers inside nanophotonic cavities to obtain strong coupling\,\shortcite{burek2012free,Hausmann2013,riedrichmoller2012one,lee2012coupling,faraon2012coupling} while maintaining their spectral stability. The optical properties of SiV centers in nanophotonic devices will be discussed in Sections\,\ref{SiVimplanted} and\,\ref{SiVnanophotonics}.  

\section{Narrow-linewidth optical emitters in diamond nanostructures via silicon ion implantation}
\label{SiVimplanted}
One major advantage of building quantum devices with solid-state emitters rather than trapped atoms or ions is that solid state systems are typically more easily integrated into nanofabricated electrical and optical structures that enable strong light-matter interactions\,\shortcite{ladd2010quantum,Vahala2003}. The scalability of these systems is important for practical realization of even simple quantum optical devices\,\shortcite{Li2015}. Motivated by these considerations and the robustness of the SiV optical transitions in bulk crystals as described in Sec.\,\ref{indistinguishable}, in Ref.\,\shortcite{evans2016narrow}, we studied the optical properties of SiV centers created in diamond nanophotonic structures via Si${}^+$ ion implantation.

Silicon-vacancy centers occur only rarely in natural diamond\,\shortcite{Lo2014}, and are typically introduced during CVD growth via deliberate doping with silane\,\shortcite{Edmonds2008,johansson2011optical} or via silicon contamination\,\shortcite{rogers2014multiple,neu2013low,clark1995silicon,sternschulte1994luminescence,Zhang2016}.
	While these techniques typically result in a narrow inhomogeneous distribution of SiV fluorescence wavelengths\,\shortcite{rogers2014multiple}, these samples have a number of disadvantages. For example, the concentration of SiV centers can be difficult to control and localization of SiV centers in three dimensions is not possible.
	Ion implantation is a commercially available technology that offers a promising solution to these problems.
	By controlling the energy, quantity, and isotopic purity of the source ions, the depth, concentration, and isotope of the resulting implanted ions can be controlled.

To create SiV centers in bulk diamond, we first implant  $\mathrm{Si}^{+}$ ions (Innovion Corporation) at a dose of $10^{10}\,\mathrm{ions}/\mathrm{cm}^2$ and an energy of 150\,keV resulting in the placement of Si atoms at an estimated depth of $100\pm20$\,nm\,\shortcite{srim}. 
After implantation, we clean the samples using an an oxidative acid clean (boiling 1\,:\,1\,:\,1 perchloric\,:\,nitric\,:\,sulfuric acid) \,\shortcite{Hauf2011} and then perform a high-vacuum ($<\,10^{-6}\,\textrm{Torr}$) anneal at a temperature of  $1100\,{}^\circ$C for two hours. 
At temperatures above $800{}^{\circ}$C, lattice vacancies can diffuse and combine with the implanted Si ions to form SiV centers. 
At higher temperatures of $1100\,{}^{\circ}$C , undesired defect complexes (e.g. divacancies) that can result in magnetic or electric field noise anneal out\,\shortcite{Acosta2009,Yamamoto2013} and partial healing of the lattice damage helps to reduce strain\,\shortcite{Orwa2011}.
	
Following these steps, SiV centers created in bulk diamond via ion implantation were characterized with photo-luminescence excitation (PLE) spectroscopy at $4\,$K.  Based on a comparison of the Si${}^+$ ion implantation density and the measured density of SiV centers in the sample,  we estimate our SiV creation yield to be 0.5--1\%. 
The PLE measurements in Ref.\,\shortcite{evans2016narrow} showed that implanted SiV centers in  bulk diamond have narrow optical transitions with linewidths of $\gamma/2\pi=320\pm180\,\mathrm{MHz}$ (mean and standard deviation for N\,=\,13 spatially resolved emitters). Almost all SiV centers had a linewidth within a factor of three of the lifetime limit. For the 13 SiV centers characterized, about half of the optical transitions were in a $15$\,GHz window. 

    \begin{figure}
    	\centering
    	\includegraphics[width=\linewidth]{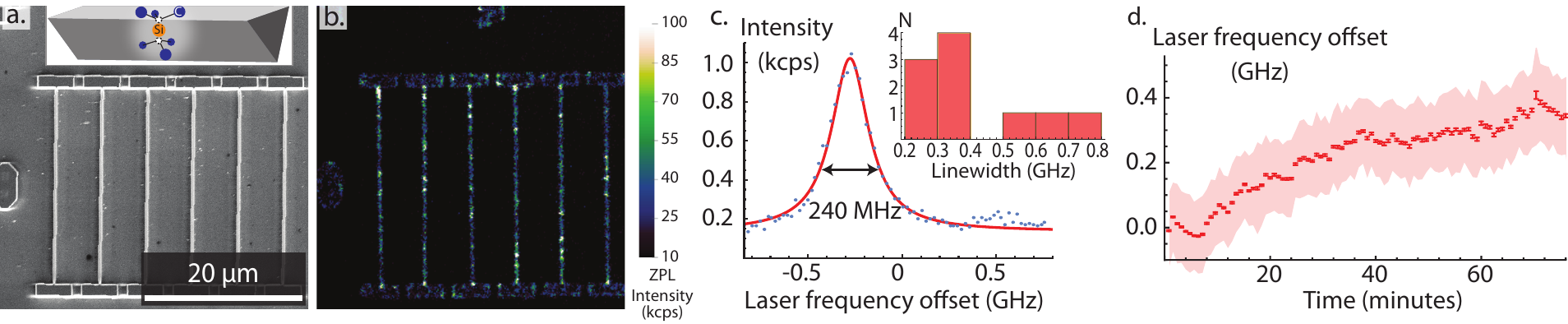}

		\caption{
			SiV centers in nanostructures.
			\textbf{(a)} Scanning electron micrograph of six nanobeam waveguides. Inset: schematic of a triangular diamond nanobeam containing an SiV center.
			\textbf{(b)} Photoluminescence image of the structures in (a). Multiple SiV centers are visible in each waveguide.
			\textbf{(c)} Linewidth of representative implanted SiV inside a nano-waveguide measured by PLE spectroscopy. 
			Inset: histogram of emitter linewidths in nanostructures. Most emitters have linewidths within a factor of four of the lifetime limit of $94$\,MHz.
			\textbf{(d)} Spectral diffusion of the emitter measured in part c. The total spectral diffusion is under $400$\,MHz even after more than an hour of continuous measurement. This diffusion is quantified by measuring the drift of the fitted center frequency of resonance fluorescence scans as a function of time. Error bars are statistical error on the fitted center position. The lighter outline is the linewidth of the fitted Lorentzian at each time point. Figure adapted from (Evans et al., 2016).
			}
	    \label{fig:implanted}
    \end{figure}
    
To test the optical coherence of the implanted SiV centers in nanophotonic devices,	we fabricated an array of diamond diamond waveguides (Fig.~\ref{fig:implanted}a) on the sample characterized above using previously reported methods\,\shortcite{burek2012free,Hausmann2013}.
	Each waveguide (Fig.~\ref{fig:implanted}a, inset) is 23\,$\mu$m long with approximately equilateral-triangle cross sections of side length 300--500\,nm.
	After fabrication, we again performed the same $1100\,{}^{\circ}$ annealing and acid cleaning procedure. This process leads to the creation of many SiV centers as seen in the  fluorescence image of the final structures of Fig.~\ref{fig:implanted}(b).

SiV centers in nanostructures display narrow-linewidth optical transitions with a full-width at half-maximum (FWHM) of $\gamma_n/2\pi=410\pm 160\,\mathrm{MHz}$ (mean and standard deviation for N\,=\,10 emitters; see Fig.~\ref{fig:implanted} inset for linewidth histogram), only a factor of $4.4$ greater than the lifetime limited linewidth $\gamma/2\pi=94\,\mathrm{MHz}$.
    The linewidths measured in nanostructures are comparable to those measured in bulk (unstructured) diamond ($\gamma_b/2\pi=320\pm 180\,\mathrm{MHz}$).
    The ratios $\gamma_n/\gamma$ and $\gamma_b/\gamma$ are much lower than the values for NV centers, where the current state of the art for typical implanted NV centers in nanostructures\shortcite{faraon2012coupling} and in bulk\,\shortcite{chu2014coherent} is $\gamma_n/\gamma>$ $100$--$200$ and $\gamma_b/\gamma>10$ ($\gamma/2\pi=13\,\mathrm{MHz}$ for NV centers).
    
    By extracting the center frequency of each individual scan, we also determine the rate of fluctuation of the ZPL frequency and therefore quantify spectral diffusion (Fig.~\ref{fig:implanted}d). Optical transition frequencies in SiV centers are stable throughout the course of our experiment, with spectral diffusion on the order of the lifetime-limited linewidth even after more than an hour.
    
The residual broadening of the optical transition can result from a combination of second-order Stark shifts and phonon-induced broadening.
The presence of a strong static electric field would result in an induced dipole that linearly couples to charge fluctuations, accounting for the slow diffusion.
Determining the precise mechanisms for the residual broadening of the SiV optical linewidths remains an important topic of future study. It should be possible to test the linear and nonlinear susceptibility of the optical transition frequencies to electric fields by applying large electric fields from nearby electrodes\,\shortcite{Acosta2012}. Despite the deviation from an ideal behavior by a factor of $4$ in linewidth and $10-100$ in ensemble inhomogenous distribution, the observed optical properties are already sufficient to observe strong atom-photon interactions in photonic crystal cavities with high cooperativity, a topic that will be discussed in the next section.  
\section{Diamond nanophotonics platform for quantum nonlinear optics }
\label{SiVnanophotonics}

\begin{figure}[t]
\begin{center}
		\includegraphics[width=\textwidth]{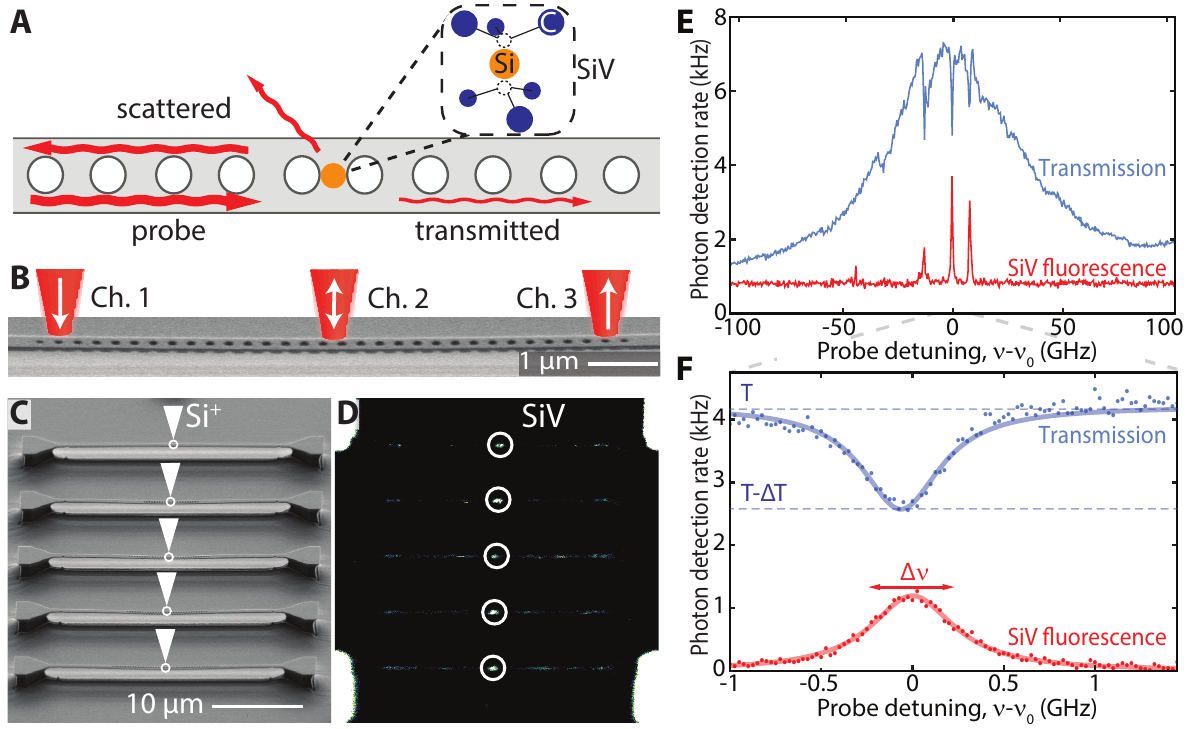}
\end{center}
		\caption{(A) Schematic of an SiV center in a diamond photonic crystal cavity. 
(B) Scanning electron micrograph (SEM) of a diamond photonic crystal cavity. 
(C) SEM of five cavities fabricated out of undoped diamond. After fabrication, SiV centers are positioned at the center of each cavity using focused Si${}^+$ ion beam implantation. 
(D) 
SiV fluorescence is detected at the center of each nanocavity shown in (C). %
(E) Measured cavity transmission and SiV fluorescence spectrum. %
	Three SiVs are coupled to the cavity and each results in suppressed transmission at the corresponding frequencies. 
(F) On resonance, a single SiV results in $ \Delta T/T=38(3)\% $ extinction of cavity transmission.  Figure reproduced from (Sipahigil et al., 2016).
}
\label{fig:cavity}
\end{figure}

We next review experiments using negatively-charged silicon-vacancy (SiV) color centers coupled to diamond nanophotonic devices that demonstrate strong interactions between single photons and a single SiV center. In Ref.~\shortcite{Sipahigil2016}, we created SiV centers inside  one-dimensional diamond waveguides and photonic-crystal cavities with small mode volumes ($V\sim\lambda^3$) and large quality factors ($Q\sim7000$) as illustrated in Fig.~\ref{fig:cavity}.
These nanophotonic devices are fabricated using angled reactive-ion etching to create free-standing single-mode structures starting from bulk diamond\,\shortcite{burek2012free,burek2014high}. A recent review article on different diamond nanophotonic device fabrication approaches can be found in Ref.\,\shortcite{Schroder16}.

To obtain optimal coupling between the emitter and the cavity mode, the emitter needs to be positioned at the field maximum of the cavity mode. This can be achieved with two possible approaches. In Refs.~\shortcite{Sipahigil2016} and \shortcite{Schroder2017}, a Si${}^+$ ion beam was focused at the center of the cavities to create SiV centers as illustrated in Fig.~\ref{fig:cavity}. Using this approach, the emitters can be positioned with close to $40$~nm precision which is limited by a combination of ion-beam spot size, alignment errors and straggle in the crystal. An alternative approach is to fabricate lithographically defined small apertures and use commerically available ion beams to implant the Si${}^+$ ions\,\shortcite{Schroder16}. Using these techniques, a large array of coupled emitter-cavity systems can be fabricated on a single sample. 

An example spectrum of the coupled SiV-cavity system at $4$\,K is shown in Fig.~\ref{fig:cavity}. Each narrow dip within the broad cavity transmission band is caused by a single SiV. On SiV resonance, the emitter results in a strong extinction ($ \Delta T/T=38(3)\% $ ) of cavity transmission\,\shortcite{Sipahigil2016}. Based on the measurements shown in Fig.~\ref{fig:cavity}F, it is possible to infer a cooperativity of $C=4g^2/\kappa \gamma= 1.0\,(1)$ for the SiV-cavity system with cavity QED parameters $\{g, \kappa, \gamma\}/2\pi=\{2.1,57,0.30\}$\,GHz where $g$ is the single-photon Rabi frequency, $\kappa$ is the cavity intensity decay rate and $\gamma$ is the SiV optical transition linewidth, cooperativity $C$ is a parameter that characterizes the ratio of coherent coupling rate ($g$) with dissipation rates ($\gamma, \kappa$). We note that the SiV optical transition linewidth of $300$\,MHz includes the sum of free-space decay, non-radiative decay and pure dephasing rates. 

\begin{figure}
\begin{center}
		\includegraphics[width=\linewidth]{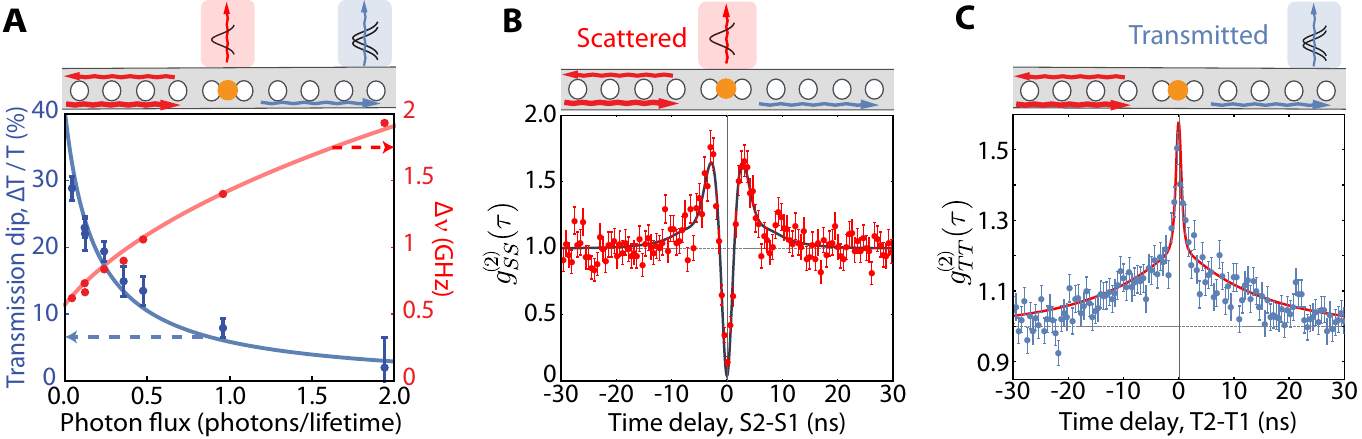}
\end{center}
		\caption{
			Single-photon nonlinearities.
	(A) Cavity transmission and SiV transition linewidth measured at different probe intensities.
	(B, C) Intensity autocorrelations of the scattered (fluorescence) and transmitted fields. 
	The scattered field shows antibunching (B), while the transmitted photons are bunched with an increased contribution from photon pairs (C).  Figure reproduced from (Sipahigil et al.,2016).
		}
	\label{fig:nonlinear}
\end{figure}

The observed cooperativity of $C=1$ in the experiments marks the onset of nonlinear effects at the single-photon level. Owing to the strong interaction between a single-photon and the SiV center, the SiV is saturated at a flux of single-photon per Purcell enhanced lifetime ($\sim300$\, ps) as shown in Fig.~\ref{fig:nonlinear}. In this regime, the strongly coupled  SiV-cavity system acts as a photon number sorter that scatters single photons while having a high transmission amplitude for photon pairs due to saturated atomic response. This effect can be observed in photon intensity correlation measurements that show antibunching for the scattered field (Fig.~\ref{fig:nonlinear}(B)) and bunching for the transmitted field (Fig.~\ref{fig:nonlinear}(C))\, \shortcite{Sipahigil2016}.

\section{Two-SiV entanglement in a nanophotonic device}
\label{twoSiVentanglement}

Experiments discussed in the previous section  demonstrated coherent single-photon single-emitter interactions. As seen in Fig.\,\ref{fig:cavity}\,(E), the optical transitions of SiV centers in nanostructures have an inhomogeneous distribution of $\sim20$\,GHz. This inhomogeneity implies that different emitters will emit spectrally distinguishable photons that are not suitable for photon-mediated interactions between multiple emitters. To generate indistinguishable photons from different nodes, it will be necessary to either improve material properties by reducing the strain variations in the crystal or develop active tuning methods. 

\begin{figure}
\begin{center}
	\includegraphics[width=\linewidth]{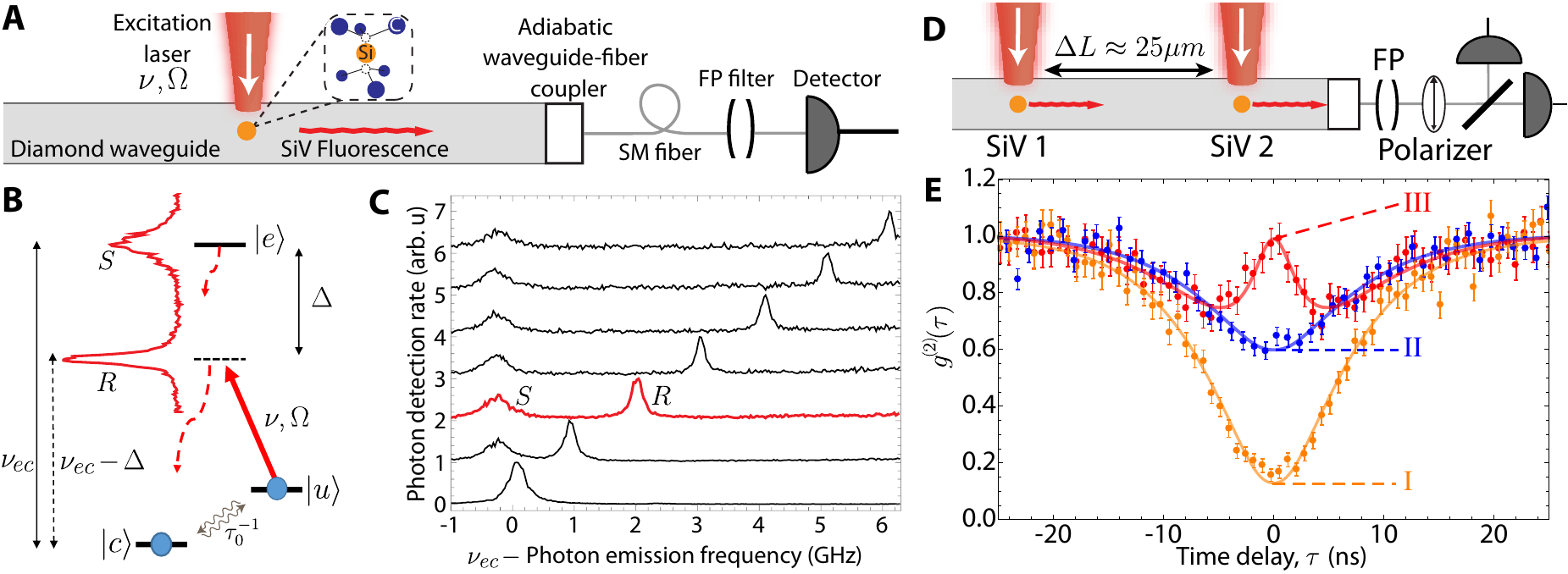}
\end{center}
		\caption{
	(A-C) Spectrally-tunable single-photons using Raman transitions.
	(A) Photons scattered by a single SiV into a diamond waveguide are coupled to a single-mode (SM) fiber.   
	A scanning Fabry-Perot (FP) cavity measures the emission spectra. 
(B) Under excitation at a detuning $\Delta$, the emission spectrum contains spontaneous emission (labeled $S$) at frequency $\nu_{ec}$ and narrow Raman emission ($R$) at frequency $\nu_{ec}-\Delta$. 
(C) $\Delta$ is varied from $0$ to $6$\,GHz in steps of $1$\,GHz and a corresponding tuning of the Raman emission frequency is observed. 
%
(D) Schematic for the two-SiV entanglement generation experiment. Photons scattered by two SiV centers into a diamond waveguide are detected after a polarizer. 
(E) Intensity autocorrelations for the waveguide photons. Exciting only a single SiV (I) yields $g^{(2)}_{single}(0)=0.16\,(3)$ for SiV1 , and $g^{(2)}_{single}(0)=0.16\,(2)$ for SiV2.
$g^{(2)}_{dist}(0)=0.63\,(3)$ when both SiVs excited and Raman photons are spectrally distinguishable (II). 
$g^{(2)}_{ind}(0)=0.98\,(5)$ when both SiVs excited and Raman photons are tuned to be indistinguishable (III).
The observed contrast between curves II and III at $g^{(2)}(0)$ is due to the collectively enhanced decay of state $|B\rangle$. Figure reproduced from (Sipahigil et al., 2016).}
	\label{fig:raman}
\end{figure}

To generate spectrally tunable photons, one can use Raman transitions between the metastable orbital states of SiV centers\,\shortcite{Sipahigil2016}. When a single SiV is excited from the state $|u\rangle$ at  a detuning $\Delta$ (Fig.\,\ref{fig:raman}), the emission spectrum includes a spontaneous component  at frequency $\nu_{ec}$ and a Raman component at frequency $\nu_{ec}-\Delta$ that is tunable by choosing $\Delta$. As shown in Fig.\,\ref{fig:raman}(C), the Raman emission frequency can be tuned by varying the excitation laser detuning $\Delta$. This technique allows frequency tuning of the Raman emission by $\pm 10$GHz which is comparable to the inhomogeneous distribution of SiV centers in nanophotonic devices.  

The Raman tuning approach enables the tuning of multiple spatially resolved emitters in a single nanophotonic device into resonance. To achieve this, we applied a separate Raman control field for two SiV centers in the same diamond waveguide\, (Fig.\,\ref{fig:raman}(D)) and measured intensity correalations of Raman photons emitted into the waveguide mode\,\shortcite{Sipahigil2016}. In this experiment, the two SiVs emit Raman photons into the same spatial mode and the frequency indistinguishability is achieved by Raman tuning. 
If the Raman emission of the two SiVs are not tuned into resonance, the photons are distinguishable, resulting in the measured $g^{(2)}_{dist}(0)=0.63\,(3)$ (curve II, Fig.~\ref{fig:raman}E) close to the conventional limit associated with two distinguishable single photon emitters $g^{(2)}_{dist}(0)=0.5$. 
Alternatively, if the Raman transitions of the two SiVs are tuned instead into resonance with each other,  an interference feature  is observed in photon correlations around zero time delay with  $g^{(2)}_{ind}(0)=0.98\,(5)$ (curve III, Fig.~\ref{fig:raman}E).

The observed enhancement of the $g^{(2)}_{ind}(0)$ results from the collectively enhanced decay of the two SiV centers. When the Raman transitions of the two SiVs are tuned into resonance with each other,
it is not possible to distinguish which of the two emitters produced a waveguide photon. Thus, the emission of an indistinguishable single photon leaves the two SiVs prepared in the entangled state $|B\rangle=(|cu\rangle + e^{i\phi}|uc\rangle)/\sqrt{2}$ \shortcite{cabrillo1999}, where $\phi$ is set by the propagation phase between emitters spaced by $\Delta L$ and the relative phase of the Raman control lasers. This state is a two-atom superradiant state with respect to the waveguide mode and scatters Raman photons at a collectively enhanced rate that is twice the scattering rate of a single emitter. This enhanced emission rate into the waveguide mode results in the experimentally observed interference peak at short time delays (curve III, Fig.~\ref{fig:raman}E) and is a signature of entanglement. 

In these experiments, the coherence time of the entangled state and control over the SiV orbital states is limited by the occupation of $\sim50$\,GHz phonons at $4$\,K, which causes relaxation between the metastable orbital states $|u\rangle$ and $|c\rangle$ and limits their coherence times to less than $50$\,ns. 
In the following Section, we discuss a microscopic model of this dephasing process and a recent experiment that extended SiV spin coherence to $13$\,ms by operation at $100$\,mK.
\section{SiV spin coherence at low temperatures}
\label{coherencesection}
In this section, we focus on the dynamics of the spin and orbital degrees of freedom of the SiV center that are relevant for storing quantum information.  In Sec.\,\ref{phonon}, we first present high temperature ($4-22$\,K) measurements of the electronic orbital relaxation dynamics  and a microscopic model of the SiV-phonon interactions that limit spin coherence~\shortcite{jahnke2015electron}. After a discussion of the predictions of this model for low temperatures, we present recent experiments that demonstrate a long coherence time of $13$\,ms at $100$\,mK\,\shortcite{Sukachev2017} in Sec.\ref{coherence}.

\subsection{Electron-phonon processes of the silicon-vacancy center in diamond}
\label{phonon}
\begin{figure}
\begin{center}
\includegraphics[width=\textwidth]{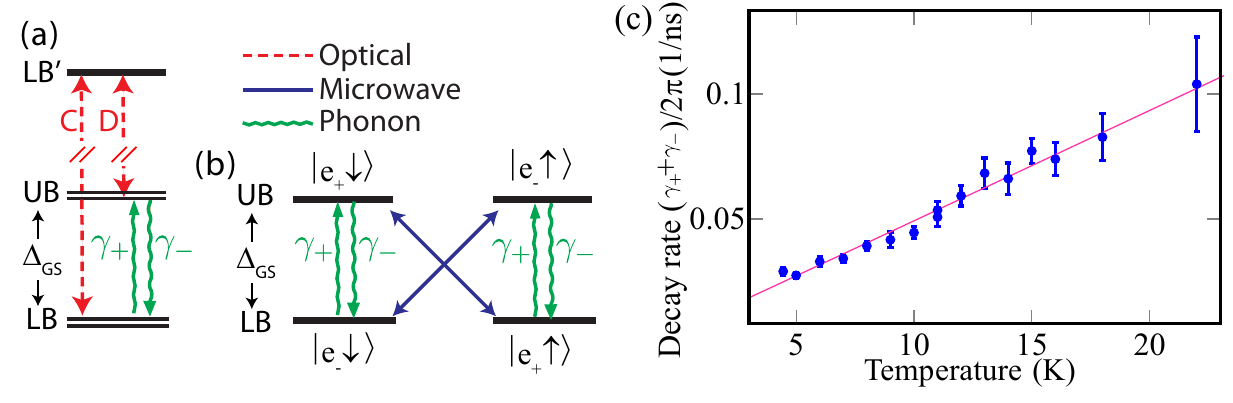}
\caption{
(a) Relevant electronic states of the SiV center at cryogenic temperatures. Optical dipole transitions (labeled C,D) connect the two spin-orbit branches UB and LB to the lower branch LB$'$ of the excited states. 
(b) Each spin orbit branch contains two electronic states with orthogonal electronic orbital and spin states (e.g. $|e_-\uparrow\rangle ,|e_+\downarrow\rangle$ for the UB).
Phonon transitions are allowed different orbital states with the same spin projection. Microwave transitions are allowed between different spin states with the same orbital projection.
(c) The measured orbital relaxation rate, $\gamma_++\gamma_-$, at different tempratures. Figure adapted from (Jahnke et al., 2015)
}
\label{fig:electronphonon}
\end{center}
\end{figure}

%
%
%
%
%
The symmetry properties of the SiV center results in ground (${}^2\mathrm{E}_\mathrm{g}$) and excited (${}^2\mathrm{E}_\mathrm{u}$) electronic orbitals that both have $E$ symmetry and double orbital degeneracy\,\shortcite{hepp2014electronic}. 
The two degenerate ground state orbitals  are occupied by a single hole with $S=1/2$ resulting in four degenerate ground states with orbital ($\{|e_+\rangle, |e_-\rangle\}$) and spin ($\{|\downarrow\rangle, |\uparrow\rangle\}$) degrees of freedom\,\shortcite{johansson2011optical,rogers2014electronic,hepp2014electronic,Gali2013} . 
The fourfold degeneracy of the ground states is partially lifted by the spin-orbit interaction ($\Delta_{GS} S_zL_z$ with $\Delta_{\textrm{GS}}\sim 45$\,GHz) which results in two ground spin-orbit branches (UB, LB in Fig.~\ref{fig:electronphonon}(a)). Each spin orbit branch consists of two degenerate states with well defined orbital and spin angular momentum (in Fig.~\ref{fig:electronphonon}(b)) at zero magnetic field\,\shortcite{hepp2014electronic}.
%
%
Acoustic phonons can also drive spin-conserving transitions between $|e_+\rangle$ and $|e_-\rangle$ orbitals, resulting in population transfer between orbitals at rates
$\gamma_{+,-}$\,\shortcite{ham1965dynamical,fischer1984vibronic}.

In Ref.~\shortcite{jahnke2015electron}, the relaxation within the ground state doublet, $\gamma_{+,-}$ in Fig. \ref{fig:electronphonon} (b), was probed directly using pulsed optical excitation and time-resolved fluorescence measurements.
This measurement was repeated for a single SiV center at various temperatures between $4.5$\,K and $22$\,K and the relaxation rate was found to scale linearly with temperature (Fig. \ref{fig:electronphonon}\,(c)).
The slowest rate observed was $(\gamma_++\gamma_-)/2\pi=(39\,\textrm{ns})^{-1}$ at $T=5$\,K. This fast decay mechanism implies that any coherence created between the states shown in Fig.~\ref{fig:electronphonon} will decay at the $100$\,ns timescale at $4$\,K. Recent experiments at $4$\,K \shortcite{pingault2014all,rogers2014all,becker2016ultrafast,pingault2017} that observed coherence times in this range can therefore be explained by this phonon-induced orbital relaxation process. 

We next discuss a microscopic model of the orbital relaxation processes within the ground electronic levels. 
The electron-phonon processes are consequences of the linear Jahn-Teller interaction between the E-symmetric electronic states and E-symmetric acoustic phonon modes\,\shortcite{fischer1984vibronic,maze2011properties,doherty2011negatively}.
%
%
Since the phonons couple to the orbital degree of freedom, phonon transitions are spin-conserving at zero magnetic field and we can focus on the orbital degree of freedom to understand the electronic dynamics of ground levels.
For a given spin state, the effective zero-field orbital Hamiltonian takes the following form
\begin{equation}
H_0=\pm\frac{1}{2}\hbar \Delta_\textrm{GS} \sigma_z\,,
\label{eq:zeroFieldHamiltonian}
\end{equation}
where $\sigma_z$ is the usual Pauli operator for orbital states in the $\{|e_+\rangle, |e_-\rangle\}$ basis, $\hbar\Delta_{\textrm{GS}}$ is the magnitude  the spin-orbit splitting, which is $-\hbar\Delta_\textrm{GS}$ for $|\uparrow\rangle$ and $+\hbar\Delta_\textrm{GS}$ for $|\downarrow\rangle$.

The interaction between the orbital states $\{|e_+\rangle, |e_-\rangle \}$ and phonon modes is described most easily if the phonon modes are linearly transformed to be circularly polarized.
With this transformation, the phonon Hamiltonian and the linear electron-phonon interaction are

\begin{align}
\mathrm{\hat{H}_E}&=\sum_{p,k}{\hbar\,\omega_k \,a^\dagger_{p,k}a_{p,k}}\\
\mathrm{\hat{V}_E}&=\sum_{k}\hbar \, \chi_k [\sigma_+ (a_{-,k}+a^\dagger_{-,k}) + \sigma_- (a_{+,k}+a^\dagger_{+,k}) ]\,,
\label{eq:electronPhonon}
\end{align}
where $\chi_k$ is the coupling strength for a single phonon, $\sigma_+$ ($\sigma_-$) is the raising (lowering) operator for the orbital states, and $a_{p,k}^\dagger$ ($a_{p,k}$) is the creation (annihilation) operator for phonons with polarization $p=\{-,+\}$ and wavevector $k$.
The coupling strength and the density of phonon modes are approximately $\overline{|\chi_k(\omega)|^2}\approx \chi\,\omega$ and $\rho(\omega)=\rho\, \omega^2$, respectively, where the overbar denotes the average over all modes with frequency $\omega_k=\omega$ and $\chi$ and $\rho$ are proportionality constants \shortcite{fu2009observation,abtew2011dynamic}.
Treating $\hat{V}_\mathrm{E}$ as a time-dependent perturbation, the first-order transitions between the orbital states involve the absorption or emission of a single phonon whose frequency is resonant with the splitting $\Delta_\textrm{GS}$ (see Fig. \ref{fig:electronphonon} (b)).
The corresponding transition rates are
\begin{align}
\gamma_+&= 2\pi \sum_{k}{ n_{-,k} |\chi_k|^2 \delta(\Delta_\textrm{GS}-\omega_{k})} \nonumber\\
\gamma_-&= 2\pi \sum_{k}{ (n_{+,k} +1)|\chi_k|^2 \delta(\Delta_\textrm{GS}-\omega_{k})}\,,
\label{eq:firstorder}
\end{align}
\noindent where $n_{p,k}$ is the occupation of the phonon mode with polarization $p$ and wavector $k$.
Assuming acoustic phonons, performing the thermal average over initial states and the sum over all final states leads to
\begin{align}
\gamma_+&= 2\pi \chi \, \rho\, \Delta_\textrm{GS}^3 n(\Delta_\textrm{GS},T) \nonumber\\
\gamma_-&= 2\pi \chi \, \rho\, \Delta_\textrm{GS}^3 [n(\Delta_\textrm{GS},T)+1]\,.
\label{eq:firstOrderRate}
\end{align}
For temperatures $T>\hbar \Delta_\textrm{GS}/k_\mathrm{B}$, Eq. (\ref{eq:firstOrderRate}) can be approximated by a single relaxation rate with a linear temperature dependence
\begin{equation}
\gamma_+ \approx \gamma_- \approx \frac{2\pi}{\hbar}\chi \rho \Delta_\textrm{GS}^2k_\mathrm{B}T\,.
\label{eq:firstOrderApprox}
\end{equation}
%
The measurements presented in Fig. \ref{fig:electronphonon} demonstrated a  linear dependence of decay rates $\gamma_{+,-}$ for temperatures below $20$\,K, but greater than the spin orbit splitting ($T>\hbar \Delta_\textrm{GS}/k_\mathrm{B}\sim2.4$\,K).
We therefore conclude that the relaxation mechanisms are dominated by a resonant single phonon process at liquid helium temperatures. For a discussion of  higher-order phonon processes that become dominant above $20$\,K, we refer the reader to Ref.~\shortcite{jahnke2015electron}.

We next discuss the implications of the linear electron-phonon interactions for qubit coherence and approaches that could be used to enhance coherence times.
%
At temperatures above the spin-orbit gap ($T>\hbar \Delta_\textrm{GS}/k_\mathrm{B}\sim2.4$\,K), any coherence formed between the states shown in \ref{fig:electronphonon}(b) will decohere quickly at the $100$\,ns timescale limited by $\gamma_{+,-}$. 
Recent experiments  that probed ground state coherences have reported $T_2^*$ values that are in good agreement with this observation\,\shortcite{pingault2014all,rogers2014all,becker2016ultrafast,pingault2017}. 
%
%
%
Eq.~\ref{eq:firstOrderRate} shows that the phonon decay rates that limit coherence are determined by a combination of phonon density of states and occupation ($\gamma_{\pm}\sim \rho(\Delta_\textrm{GS})(2\, n(\Delta_\textrm{GS},T)+1\mp1)$) at the energy of the spin-orbit splitting with $\Delta_\textrm{GS}\sim45$GHz.
Since the interaction with the phonon bath is a Markovian process, dynamical decoupling sequences cannot be applied to extend coherences.
To extend $T_2^*$, we will therefore focus on approaches that reduce the orbital relaxation rates $\gamma_\pm$.
The first two approaches focus on reducing phonon occupation $n(\Delta_\textrm{GS},T)$ to decrease $\gamma_+$.
The occupation depends on the ratio, $T/\Delta_{\textrm{GS}}$ , of the temperature and the energy splitting between the coupled orbital states.
Substantial improvements can be achieved by minimizing this ratio in cooling the sample to lower temperatures ($T \ll \Delta \sim 2.4$\,K).
Based on the fit in Fig.~\ref{fig:electronphonon}, the expected orbital relaxation timescale is given by $1/\gamma_+=\,200\,(e^{2.4\textrm{K}/T}\,-1)\,$\,ns which correspond to $2~\mu$s at $1$\,K and $2$\,ms at $0.26$\,K. A recent experiment by \shortcite{Sukachev2017} operating at these low temperatures will be discussed in Sec.~\ref{coherence}.
A second approach is to increase $\Delta_\textrm{GS}$ by using emitters subject to high strain that increases the splitting between the spin-orbit branches.
At the limit of $\hbar \Delta_\textrm{GS} \gg kT$, similar reductions in phonon occupation can be used to suppress relaxation rates at $4$\,K, an effect that was recently observed in Ref.\,\shortcite{Sohn2017}. 
In both cases, only the two lowest energy states constitute a subspace that does not couple to phonons.
The lowest two energy states are therefore expected to have long coherence times and could be used as a long-lived spin qubit.

We note that the linear electron-phonon interaction Hamiltonian (Eq.~\ref{eq:electronPhonon}) and the resulting single-phonon orbital relaxation process are analogous to the Jaynes-Cummings Hamiltonian and Wigner-Weisskopf model of spontaneous emission used in quantum optics. 
One can therefore use ideas developed in the context of cavity QED to minimize relaxation rates $\gamma_{+,-}$ by use of phononic bandgaps\,\shortcite{Burek16} or strong coupling to well defined nanomechanical modes to realize multi-qubit interactions\,\shortcite{Burek16,Sohn2017}. 
%
%
%
\subsection{SiV spin at low temperatures: a long-lived quantum memory}
\label{coherence}
%
%
\begin{figure}
	\includegraphics[width=\linewidth]{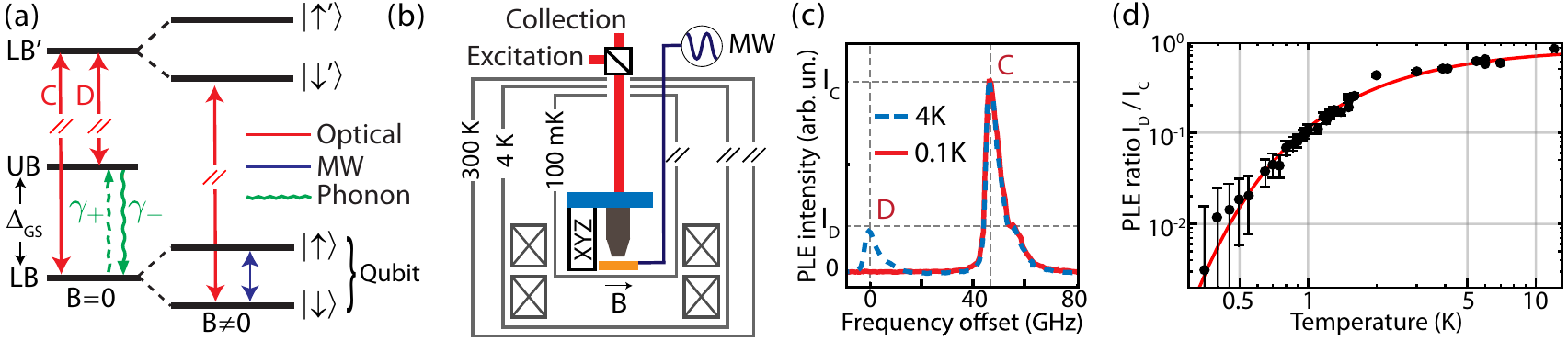}
		\caption{
		(a) SiV electronic structure. Optical transitions C and D connect the lower (LB) and upper (UB) spin-orbit branches to the lowest-energy optical excited state $\left(\text{LB}'\right)$. 
	Each branch is split into two spin sublevels in a magnetic field $\vec{B}$. 
	$\gamma_+$ and $\gamma_-$ are phonon-induced decay rates.  
	 (b) Schematic of the setup. An objective is mounted on piezo positioners to image the diamond sample using free-space optics.  The combined system is attached to the mixing plate of a dilution refrigerator and placed inside a superconducting vector magnet. 
	(c) PLE spectra of an SiV ensemble at $B = 0$ for $T=4$\,K and $0.1$\,K. The peak intensity $I_C$ ($I_D$) is proportional to the population in the LB (UB). (d) 
	$I_D / I_C$ (and $\gamma_+/\gamma_-$) is reduced at low temperatures, following $e^{-\hbar\Delta / k_B T}$ with $\Delta_{\textrm{fit}}=42 \pm 2$\,GHz in agreement with the measured $\Delta_{\textrm{GS}}=48$\,GHz. Figure adapted from (Sukachev et al., 2017)
	}
\label{fig:lowtemperature}
\end{figure}

Motivated by the prediction of long spin coherence at low temperatures, we probed the coherence properties of single SiV centers at mK temperatures in Ref.~\shortcite{Sukachev2017}. The key idea of  this experiment can be understood by considering the energy level diagram of the SiV in Fig.~\ref{fig:lowtemperature}(a). Application of a magnetic field lifts the degeneracy between the spin-orbit states in the LB ($|e_+\uparrow\rangle$ and $|e_-\downarrow\rangle$ abbreviated as $|\uparrow\rangle$ and $|\downarrow\rangle$ from here onwards) with different spin projections. These states, which are defined as qubit states, decay at a rate $\gamma_+$ determined by the phonon occupation at frequency $\Delta_{\textrm{GS}}$~(Eq.~\ref{eq:firstOrderRate}). 
%
By reducing the occupation of phonon modes at $\Delta_{\textrm{GS}}$ at lower temperatures, one can  suppress the rate $\gamma_+$, leaving the spin qubit in a manifold free from phonon-induced decoherence, thereby increasing spin coherence ~\shortcite{jahnke2015electron}.

In Ref.~\shortcite{Sukachev2017}, we investigated the SiV spin properties below $500$\,mK using a dilution refrigerator with a free-space confocal microscope and a vector magnet as shown in Fig.~\ref{fig:lowtemperature}(b). 
We first studied the thermal population of the LB and the UB between $0.1$ and $10$\,K using an ensemble of as-grown SiV centers to probe the effective sample temperature. 
We probed the relative populations in the LB and the UB (and therefore the ratio $\gamma_+/\gamma_-$ ) by measuring the absorption spectrum of transitions C and D. Transitions C and D were both visible in PLE at 4\,K, which indicates comparable thermal population in the LB and UB [Fig.~1(c)]. 
As the temperature was lowered [Fig.1(d)], the ratio of the transition D and C peak amplitudes ($I_D / I_C$) reduces by more than two orders of magnitude and follows $e^{-\hbar \Delta_\text{GS} / k_B T}$~\shortcite{jahnke2015electron}. These measurements demonstrated an orbital polarization in the LB of $> 99\%$ below $500$\,mK. At these low temperatures, $\gamma_+ << \gamma_-$ and the qubit states are effectively decoupled from the phonon bath. 
%


\begin{figure}
	\includegraphics[width=\linewidth]{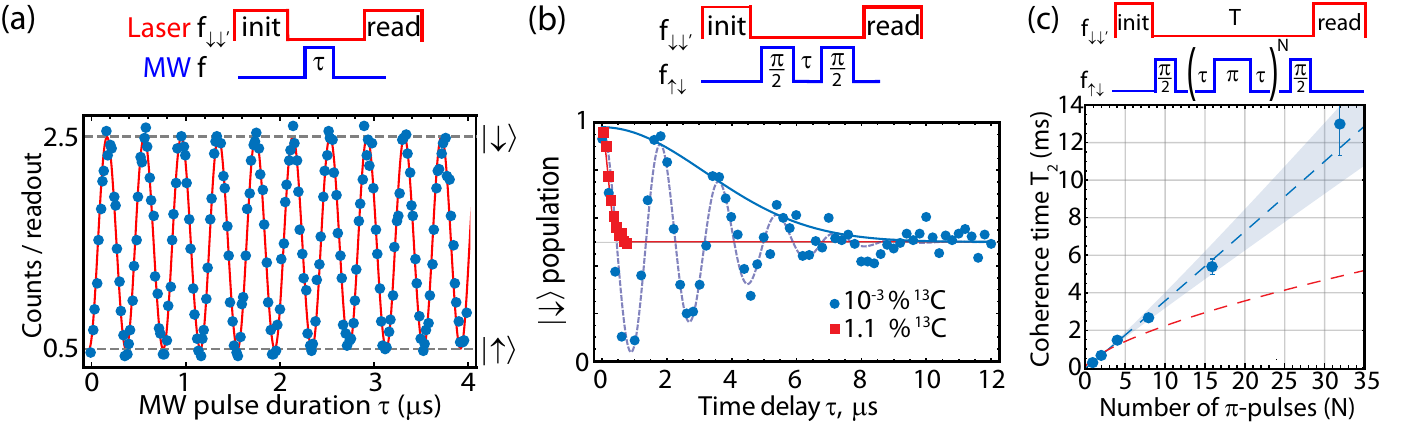}
		\caption{
	Coherent spin control. (a) Resonant driving at frequency f$_{\uparrow\downarrow}$ results in Rabi oscillations between states $|\uparrow\rangle$ and $|\downarrow\rangle$. 
	(b) Ramsey interference measurement of $T_2^*$ for the two samples (see text for details). MW pulses are detuned by $\sim 550$\,kHz from the $f_{\uparrow \downarrow }$ for the blue circles.  
(c) $T_2$ coherence vs. number of rephasing pulses $N$ for Sample-12. Fitting to $T_2 \propto N^\beta$ gives $\beta = 1.02 \pm 0.05$ (blue dashed line), the shaded region represents a standard deviation of 0.05. For comparison, the red dashed line shows $N^{2/3}$ scaling. 
Figure adapted from (Sukachev et al., 2017).	}
	\label{fig:coherence}
\end{figure}
We used microwave fields resonant with the qubit transition to coherently control the SiV spin and probe its coherence\,\shortcite{Sukachev2017}. 
In these experiments, single strained SiV centers with $\Delta_\text{GS} \sim 2\pi \,80$\,GHz are used. When crystal strain is comparable to spin-orbit coupling ($\sim 48$\,GHz), the orbital components of the qubit states are 
no longer orthogonal~\shortcite{hepp2014electronic}, leading to an allowed magnetic dipole transition between the qubit states~\shortcite{pingault2017}.
Fig.~\ref{fig:coherence} shows measurements where a resonant microwave field drives coherent Rabi oscillations of the spin qubit. In these experiments, a long laser pulse at frequency f$_{\downarrow\downarrow'}$ first initializes the spin in state $|\uparrow\rangle$ via optical pumping. 
After a microwave pulse of duration $\tau$, a second laser pulse at f$_{\downarrow\downarrow'}$ reads out the population in state $|\downarrow\rangle$.
Ramsey interference was used to measure the spin dephasing time $T_2^*$ for SiV centers in two different samples [Fig.~\ref{fig:coherence}(b)]. 
For a sample that contained a low concentration of ${}^{13}$C nuclear spins (Sample-12, blue circles), we measured a dephasing time in the range of $T_2^* \sim 4\,\mu$s. 
For a second sample that contained a natural abundance of ${}^{13}$C  nuclear spins (Sample-13, red squares), we measured $T_2^* \approx 300$\,ns
which is similar to typical values observed with NV centers. 
These results demonstrate that the dephasing time $T_2^*$ of SiV centers is primarily limited by the nuclear spin bath in the diamond host with a natural abundance of ${}^{13}$C~\shortcite{Childress2006a}. 

Dephasing due to slowly evolving fluctuations in the environment (e.g. nuclear spins) can be suppressed by using dynamical decoupling techniques~\shortcite{Ryan2010,DeLange2010}.
We extended the spin coherence time $T_2$ by implementing Carr--Purcell--Meiboom--Gill (CPMG) sequences with $N=1,2,4,8,16$, and 32 rephasing pulses~\shortcite{Meiboom1958} in the isotopically purified sample. 
Fig.~\ref{fig:coherence}(c) shows that the coherence time  increases approximately linearly with the number of rephasing $\pi-$pulses $N$.
The longest observed coherence time is $T_2=13\pm 1.7$\,ms for $N=32$. 
Repeating the CPMG sequences for $N=1,2$, and 4 with the Sample-13 gave similar coherence times $T_2$ as for Sample-12. Surprisingly,  the observation that the coherence time $T_2$ in both samples is identical for a given $N$ indicates that the coherence time $T_2$ is not limited by the nuclear spin bath, but by another noise source. While the origin of the noise source is at present not understood, the linear dependence of $T_2$ on $N$ suggests that  $T_2$ can potentially be further improved by using additional rephasing pulses~\shortcite{Sukachev2017}. 

These observations establish the SiV center as a promising solid-state quantum emitter
for the realization of quantum network nodes using integrated diamond nanophotonics~\shortcite{Sipahigil2016}. 
The demonstrated coherence time of 13\,ms is already sufficient to maintain quantum states between 
quantum repeater nodes separated by $10^3$\,km~\shortcite{Childress2006a}. 
The quantum memory lifetime could be further extended by implementing robust dynamical decoupling 
schemes~\shortcite{DeLange2010} or using coherently coupled nuclear spins as longer-lived memories~\shortcite{maurer2012room}.

\section{Outlook}
\label{Outlook}

In this Chapter, we reviewed recent progress on the realization of near-deterministic spin-photon interactions based on a single SiV in a cavity and two-emitter entanglement generation in a single nanophotonic device\,\shortcite{Sipahigil2016}. When combined with the recent demonstration of 13ms spin coherence for SiV centers \,\shortcite{Sukachev2017}, these advances make SiV centers the first solid-state quantum emitter that has the desired combination of long-lived spin coherence with a deterministic spin-photon interface. 

Here we outline several research directions to advance this diamond nanophotonics platform towards the realization of scalable quantum repeater nodes. Our preliminary results with improved cavity designs and emitter positioning yield higher cooperativities of $C\sim10$ . These advances open up the possibility of realizing deterministic entanglement generation between two-SiVs in a single nanocavity at high rates while maintaining long spin coherence at low temperatures. To achieve this, current experiments are focusing on combining coherent spin control methods\,\shortcite{Sukachev2017,Becker2017} and efficient fiber collection from nanophotonic structures\,\shortcite{burek2017} at 100mK. For long-distance quantum communication applications, these devices are expected to improve the entanglement generation rates by many orders of magnitude compared with the state-of-the-art \shortcite{Rosenfeld2017,hucul2015modular,hensen2015loophole}. We expect that these technical improvements will bring the realization of a multi-node quantum repeater within reach. For long-distance quantum communication, the SiV photons should  be downconverted to the telecommunication-band for low-loss photon distribution in a fiber network. The improved collection efficiency from the nanophotonic sturctures is expected to result in a high signal-to-noise ratio quantum frequency conversion to the $1310-1350$\,nm telecommunication band with reduced spontaneous noise contribution from a pump in the $1620-1685$\,nm band \shortcite{Zaske2012}. 

Finally, other color centers such as the germanium-vacancy (GeV)\,\shortcite{palyanov2015germanium} and the tin-vacancy (SnV)\,\shortcite{iwasaki2017tin} centers with similar symmetry properties could yield improved optical and spin properties and recently started being actively explored. The GeV center was demonstrated to be superior to SiV centers in terms of its quantum efficiency\,\shortcite{bhaskar2017} and is expected to show similar spin properties at dilution fridge temperatures\,\shortcite{Siyushev2017}. To date, however, there has not been a system that combined good optical and spin properties at higher temperatures of $4$\,K. In principle, no fundamental reason prevents obtaining spectrally stable optical transitions and long spin coherence times at $4$\,K. Recent experiments with the neutral silicon-vacancy centers have already started investigating this possibility\,\shortcite{green2017,rose2017observation}. Considering the existence of hundreds of color centers in diamond\,\shortcite{Zaitsev2001}, a systematic experimental survey of different color centers with guidance from density functional theory calculations\, \shortcite{goss2005vacancy} is likely to result in new systems with improved device performance at elevated temperatures, which could be instrumental for realizing practical solid-state quantum repeater nodes. 
\section{Acknowledgements}
The work described in this Chapter highlights the results of a collaborative team effort over the past decade. Most of the experiments presented were carried out by members of the Harvard quantum optics group including Ruffin Evans, Denis Sukachev, Mihir Bhaskar, Christian Nguyen and Alexander Zibrov. Furthermore, 
we are grateful to Fedor Jelezko and members of the Ulm quantum optics group, including Kay Jahnke, Lachlan Rogers, Mathias Metsch and Petr Siyushev for a close collaboration on quantum optics with novel color centers; to Marko Loncar, Hongkun Park and Michael Burek for critical contributions including the development and fabrication of the diamond nanophotonic devices. We additionally acknowledge many discussions and collaborations with  Ed Bielejec, Ryan Camacho, and Dirk Englund. This work was supported by the NSF, CUA, AFOSR MURI, ARL CDQI and DURIP Grant No. N00014-15-1-28461234 through ARO.

\bibliographystyle{OUPnamed}
\bibliography{SiVbib}
\end{document}